# Orbital angular momentum dichroism caused by the interaction of electric-magnetic dipole moments in chiral metal nanoparticles

Yangzhe Guo[1], Guodong Zhu[1], Wanlu Bian[2], Bin Dong[3,*], and Yurui Fang[1,*]

[1.] Key Laboratory of Materials Modification by Laser, Electron, and Ion Beams (Ministry of Education); School of Physics, Dalian University of Technology, Dalian 116024, P.R. China.

[2.] School of Physics, Liaoning University, Shenyang 110036, P.R. China

[3.] Key Laboratory of New Energy and Rare Earth Resource Utilization of State Ethnic Affairs Commission, Key Laboratory of Photosensitive Materials & Devices of Liaoning Province, School of Physics and Materials Engineering, Dalian Nationalities University, 18 Liaohe West Road, Dalian 116600, PR China

*Corresponding authors: Bin Dong, dong@dlnt.edu.cn; Yurui Fang, yrfang@dlut.edu.cn

**Abstract**

Circular dichroism (CD) caused by the response of a chiral object to circularly polarized light has been well established, and the strong CD of plasmonic meta-molecules has also become of interest in recent years; however, their response if the light also has orbital angular momentum is unclear. In this paper, the dichroism of a plasmonic cuboid-protuberance chiral structure under the illumination of a light beam with both orbital and spin angular momentums is numerically investigated. Distinguished spectra are observed under the different momentums. The circular dichroism under the combination of vortex beam and light spin is enhanced. This phenomenon is attributed to the partial spatial excitation of the nanoparticle, and the strong dichroism is simultaneously caused because of the interaction of the induced electric and magnetic modes and other higher-order modes caused by the partial excitation of the vortex beam. This research provides further insight into chiral light-matter interactions and the dichroism of light with orbital angular momentum.

**Introduction**

　　It is well known that light can have both spin angular momentum (SAM) and orbital angular momentum (OAM). The different responses of chiral matter to circularly polarized light (CPL) have been well established. In particular, in recent years, plasmonic chirality has become a popular topic and has attracted substantial research attention. Chiral molecules in nature have quite weak scattering optical activity (OA) cross-sections, while plasmonic structures can enhance the circular dichroism (CD) because of the significant enhancement of both of the electromagnetic field[1-3] and super chiral field[4-6] in the vicinity of the structures.[7] Moreover, plasmonic structures with chiral morphologies (which are also called chiral meta-molecules) are designed for strong OA response. Strong CD responses have been recently observed in various structures, such as swastikas,[8-10] DNA-based assembled gold particles,[11-13] crossed rods,[14,15] pairs of mutually-twisted planar metal patterns,[16] metal helices,[17-20] and oligomers of nanodisks,[21-23] among others.[24-27] The mechanism of the CD response of chiral plasmonic structures has been proposed by different models. The simplest model is the Born-Kuhn model based on a phenomenological explanation, in which the $x$ and $y$ components of the electric field vectors with different dilations overlap with the two spatially separated rods at advanced/delayed times to cause varied responses[15]. For planar symmetric structures under oblique excited light, the CD is explained as the coupling of the electric and magnetic modes.[28] Some other works have also explained the CD from the perspective of the symmetry of the structures[27]. A more profound explanation is the analogy between plasmonic meta-molecules and chiral molecules, the chiral response mechanism of which has been sufficiently addressed.[29] The CD response of chiral molecules originates from the coupling of electric-magnetic dipole moments and the coupling of electric dipole-quadrupole moments. Electric-magnetic dipole coupling is usually brought in in plasmonic chirality and can be described as the mixed electric and magnetic polarizability, while the coupling of electric dipole-quadrupole moments is ignored because it is weaker. The OAM of light beams established by Allen et al[30] became characteristic of light beams, and has been used for the classification of light,[31] information encoding,[32,33] optical trapping rotation,[34] quantum entanglement,[35,36] super-resolved imaging,[37] the photonic spin Hall effect,[38] spinning object detection,[39] and wavefront manipulation,[40] among others.[41-44] Additionally, the OAM has analogous effects to those of SAM during light-matter interactions, such as CD. Although the dichroism induced by circular polarization has been extensively studied, there have been fewer reports on the dichroism of OAM light. Recently, Kerber et al. employed nanoantennas to detect the status of a twisted light beam via phase conversion.[45,46]. They also extended the scattering dichroism of OAM light of nanoparticles[47]; however, in this work, they did not propose a detailed mechanism that accounts for the large difference.

In the present work, a simple cuboid-protuberance chiral structure under a Laguerre-Gaussian beam with both SAM and OAM was investigated via finite element method numerical simulations. The dichroisms of different OAM and SAM combinations were studied, and the OAM beam-induced electric-magnetic mode interactions were used to explain the strong dichroism. The mixed electric and magnetic polarizability was used to deduce the extinction of the structure under a vortex beam. Under different OAM excitation modes, the extinction spectra were found to be quite different and resonant at different wavelengths. The results demonstrate that OAM primarily results in the higher-order oscillations of the particle because of the partial excitation, and SAM primarily results in an optical chiral field.

**Simulation model**

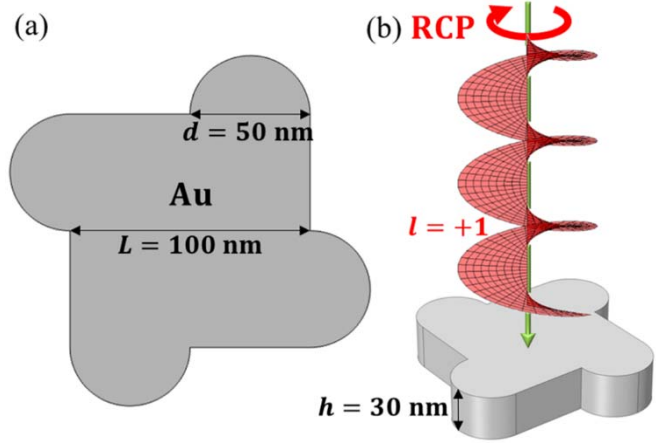

Figure 1. (a) Top view of a chiral gold nanostructure. (b) Schematic diagram of chiral gold nanostructures vertically excited by L-G beams with $l = +1$ and $s = +1$.

In the Cartesian coordinate system, the electric field distribution of a Laguerre-Gaussian (L-G) beam propagating along the $z$-axis is divided into three components. The three components of the harmonic varying electric field amplitude (the time factor $e^{i\omega t}$ is hidden in the following because of harmonicity) in the paraxial approximation can respectively be expressed as

$$\tilde{E}_{bx} = \frac{1}{\sqrt{2}} E_0 \cdot F \cdot e^{i\phi}, \tag{1a}$$

$$\tilde{E}_{by} = \frac{1}{\sqrt{2}} E_0 \cdot F \cdot e^{i\phi} \cdot e^{i\Delta\varphi}, \tag{1b}$$

$$\tilde{E}_{bz} = 0, \tag{1c}$$

the amplitude and phase of which are as follows:

$$F = \sqrt{\frac{\pi}{2}} C(p,l) \frac{w_0}{w(z)} \left(\frac{r\sqrt{2}}{w(z)}\right)^{|l|} L_p^{|l|}\left(\frac{2r^2}{w^2(z)}\right) \exp\left(-\frac{r^2}{w^2(z)}\right), \tag{2}$$

$$e^{i\phi} = \exp\left(-ik\frac{r^2}{2R(z)}\right) \exp(-ik(z-z_0)) \exp(-il\varphi) \exp(i\psi(z)), \tag{3}$$

where $E_0 = 1 V/m$, $\Delta\varphi = \frac{\pi}{2} s$, $s = \pm 1, 0$ represents right / left (RCP/LCP) circularly polarized and linearly polarized light, respectively, $r = \sqrt{(x-x_0)^2 + (y-y_0)^2}$ is the radial distance from the center axis of the beam, and $z$ is the axial distance from the beam's focus. Additionally, $C(p,l) = \sqrt{\frac{2p!}{\pi(p+|l|)!}}$, $l$ represents different OAMs, $L_p^{|l|} = L_p^{|l|}(p, |l|, 2\frac{r^2}{w^2(z)})$ is the generalized Laguerre polynomials (for simplicity, we consider the single-ring LG beams with zero radial index $p = 0$), and $w(z)$ is the beam radius at position $z$ and is defined as $w(z) = w_0 \sqrt{1 + (\frac{z-z_0}{z_R})^2}$, $z_0 = 0\ nm$. The beam waist $w_0 = w(0) = 200\ nm$ is fixed to get tightly focused beam in all wavelengths. The paraxial approximation condition is effective below 800 nm. And above 800 nm condition will be discussed in the discussion part. Finally, $z_R = \frac{kw_0^2}{2}$ is the Rayleigh range, which is related to the waist at the focus, in which $k = 2\pi\frac{n}{\lambda_0}$, where $\lambda_0$ is the wavelength of the light in a vacuum and $n = 1.25$ is the effective refractive index of the surrounding medium. In Eq. (3), $R(z) = z(1 + (\frac{z_R}{z})^2)$ is the radius of curvature of the beam's wavefronts at $z$ and $\psi(z) = (2p + |l| + 1)\arctan(\frac{z}{z_R})$ is the Gouy phase at $z$. The phasor sign is negative just follow the definition of the software.

The OAM is quantified with $l = 0, \pm 1, \pm 2, ...$, and SAM is quantified with $s = 0, \pm 1$. In Figure 1a, a cuboid-protuberance gold chiral nanostructure is shown to consist of a block with a side length of $L = 100\ nm$ and four hemispheres with diameters of $d = 50\ nm$. The thickness $h$ is 30 nm. The nanostructure is placed in the $z_0$ position and excited by an L-G beam with both OAM and SAM in the normal direction, as shown in Figure 1b.

In the study, the dichroism $\Delta\sigma$ is defined as
$$\Delta\sigma_{l,s=l1,s1;\ l2,s2} = \sigma_{l1,s1} - \sigma_{l2,s2}, \tag{4}$$
where $\sigma_{l,s}$ represents the extinction cross-section of the particle under the illumination of an L-G light with OAM $l$ and SAM $s$. When $l = l_1$ is fixed, $\Delta\sigma_{l,s=l1,s1;\ l1,s2}$ represents the SAM CD $\Delta\sigma_{l1}^{SAM}$ (and a traditional CD for $l = 0$). When $s = s_1$ is fixed, $\Delta\sigma_{l,s=l1,s1;\ l2,s1}$ represents the OAM CD $\Delta\sigma_{s1}^{OAM}$. When the total angular momentum (TAM) $j = |l + s|$ is fixed, $\Delta\sigma_{l,s=l1,s1;\ l1,s2}$ represents the total angular momentum CD $\Delta\sigma_{l,s=+j;\ -j}^{TAM}$.

Model simulations were performed with full-wave numerical simulations by using a commercial finite element method software package (COMSOL Multiphysics 5.4). The cuboid-protuberance nanostructure was placed in a homogeneous medium with $n = 1.25$. Non-uniform meshes (the largest mesh was less than $\lambda/6$ fixed at $\lambda = 500\ nm$ and the smallest mesh was about 2 $nm$) were used to format the object. Perfectly matched layers (PML) were used to minimize the scattering from the outer boundary. The scattering cross-sections were obtained by integrating the scattered power flux over an enclosed surface outside the structure, and the absorption cross-sections were obtained by integrating the ohmic heating within the structure. The dichroism was calculated as defined previously.

## Results and discussion
### Combined dichroism
Figure 2a presents the extinction cross-section spectra of the structure under the excitation of an L-G beam with $l = 0, \pm 1$ and $s = 0, \pm 1$. The spectra of $l = 0$ are the same as the normal resonances of a particle without a vortex (data not shown) with the main resonant peak at around 810 nm and two smaller peaks. An obvious difference of the spectra emerged when OAM ($l \neq 0$) was added, and the resonance peaks were quite different (peaks 4, 5, 6, 7, and 8 in Figure 2a). For $l = +1, s = +1$, $l = -1, s = -1$, and $l = \pm 1, s = 0$, the main resonant peaks were red-shifted to around 1000 nm as compared with the no-vortex excitation. For $j = 0$, the main resonant peaks were blue-shifted to around 500 nm and a broad peak at 1470 nm appeared. It can also be found from the spectra that the total extinction cross-section decreased with the decrease of $|j|$ ($l = 1$) from 2 to 0. Interestingly, every two spectra respectively distinguished by a line and symbol almost overlap with each other: $l = +1, s = +1$ coincides with $l = -1, s = -1$, $l = +1, s = 0$ coincides with $l = -1, s = 0$, and $l = +1, s = -1$ coincides with $l = -1, s = +1$. Additionally, the spectra exhibited little differences for the same values of $|j|$ and $|l|$. Figure 2b presents the dichroism of the almost-overlapped spectra. For the normal CD ($l = 0$), the maximum value of $\Delta\sigma_{l,s=0,+1; 0,-1}$ is obtained at the maximum peak 3 of the extinction spectrum, and a negative value at the peak 1 is that of a normal chiral phenomenon. For $|l| = +1$, the main CD peak at around 550 nm decreases with the decrease of $|j|$. Figures 2c-2f present the electric field intensity distributions and surface charge distributions of the system at the peaks shown in Figure 2a. Figure 2c presents the $l = 0$ excitation, which is the same as the traditional CPL excitation with dipole and higher-order modes. For $|j| = 2$, the resonance mode (990 nm) is similar to a whole-surface mode oscillating vertically (Figure 2d), the surface charge density of which is less than that of the $l = 0$ excitation. For $|j| = 1$ (Figure 2e), the main peak at 1020 nm similarly presents a whole-surface mode. The small peak at 700 nm is a higher-order mode with long, saddle-shaped strip charges that cross and alternate with each other. The peak at 540 nm is a higher-order mode. For $|j| = 0$ ($|l| = +1$, Figure 2f), the peak at around 1470 nm shows a rotational crossed charge that lowers the energy. The peak at 700 nm is very similar to that of the $|j| = 1$ condition, and the peak at 500 nm is a higher-order mode. The reason for the formation of such patterns when $|l| = 1$ is that the vortex beam wavefront arrives at different nanoparticle parts at different times, and the polarization alternates simultaneously. Therefore, because of the special beam wavefront of the vortex light, the combination of the OAM and SAM of the light will lead to diverse optical responses of the chiral structures.

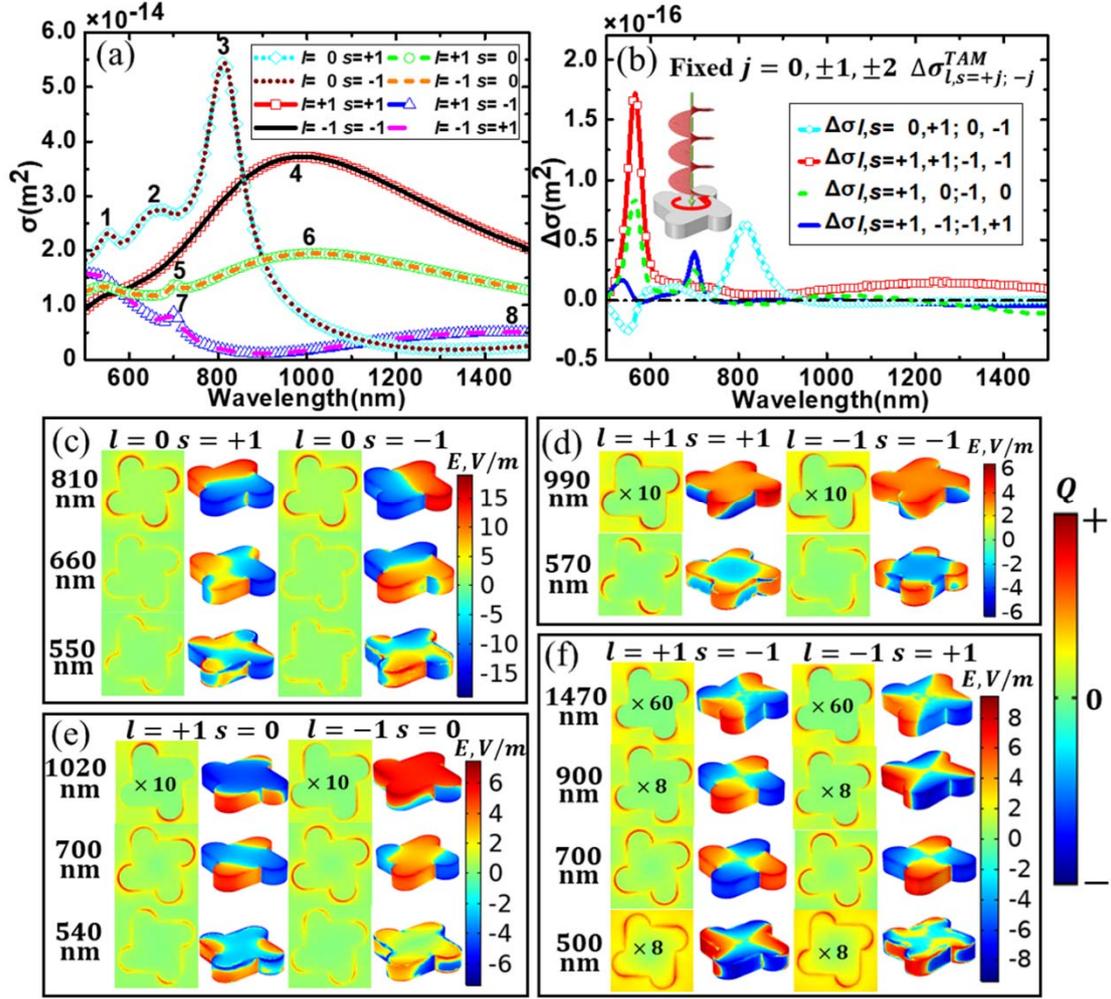

Figure 2. (a) Extinction cross-section spectra of a chiral gold structure under incident light with $l = 0, \pm 1$ and $s = 0, \pm 1$. (b) Corresponding four dichroisms. The cyan diamond solid line represents normal CD, and the red square solid, green short dash, and blue solid lines represent $\Delta\sigma_{l,s=+j;\,-j}^{TAM}$ for $|j| = 2, 1,$ and $0,$ respectively. (c-f) Corresponding electric field intensity (column 1 and 3 for each subpanel) and surface charge distributions (column 2 and 4 for each subpanel) at the selected locations.

In additional to dichroism described previously, other combinations also can be derived by Eq. (4). Figure 3a presents the dichroism $\Delta\sigma_{l1}^{SAM}$ for a fixed OAM of light but with an LCP or RCP SAM. Excluding the condition of $l = 0$ (the black solid line) discussed previously, the other two conditions result in quite different responses, the dichroisms of which are almost two orders of magnitude larger than the traditional dichroism. When the orbital momentum is reversed, the $\Delta\sigma_{l1}^{SAM}$ spectrum is also reversed, and their profiles are symmetric. Figure 3b presents the dichroism $\Delta\sigma_{s1}^{OAM}$ for a fixed SAM of light but with different OAMs. The $s = 0$ condition was presented previously. For the conditions of $s = \pm 1$, $\Delta\sigma_{s1}^{OAM}$ is also reversed but symmetric when the spin momentum changes its sign. The dichroisms $\Delta\sigma_{s1}^{OAM}$ for $s = \pm 1$ are also two orders of magnitude larger than that of the traditional dichroism. Therefore, the combination of SAM and OAM can yield a much greater CD, which may be useful for nano-sized chiral device design applications.

The conditions discussed previously all have symmetric excitation beam conditions, such as $\pm|l|$, $\pm|s|$, and $\pm|j|$. However, there remain other combinations of $l$ and s, such as $\Delta\sigma_{l,s=+1,+1;+1,0}$ and $\Delta\sigma_{l,s=+1,+1;-1,0}$, etc. (see Figure 3c-f),, which all have quite strong dichroisms.

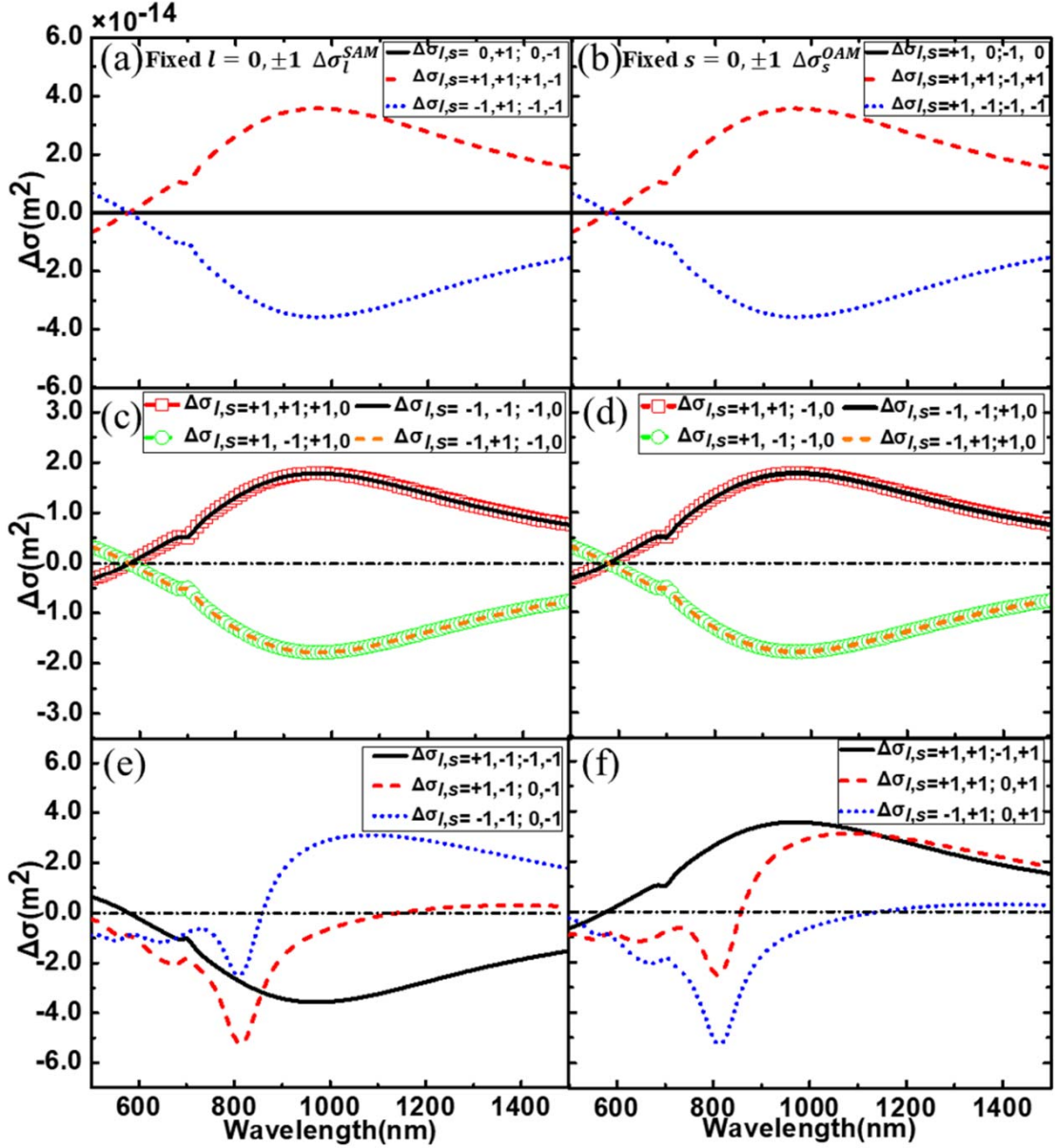

Figure 3. (a) The dichroism of $\Delta\sigma_{l,s=l,+1;l,-1}$ when $l = 0, \pm 1$. (b) The dichroism of $\Delta\sigma_{l,s=+1,s;-1,s}$ when $s = 0, \pm 1$. (c) The dichroism of $\Delta\sigma_{l,s=l,s;l,0}$ when $l = \pm 1$, $s = \pm 1$. (d) The dichroism of $\Delta\sigma_{l,s=l,s;-l,0}$ when $l = \pm 1$, $s = \pm 1$. (e-f) The dichroism of $\Delta\sigma_{l,s=+1,s;-1,s}$, $\Delta\sigma_{l,s=+1,s;0,s}$, and $\Delta\sigma_{l,s=-1,s;0,s}$ when $s = \pm 1$.

To understand the connection between OAM and SAM in the dichroism illustrated previously, the coupled electric-magnetic dipole model was investigated with similar steps as those conducted by Tang,[4] and the optical chirality C is given by the following formula.

$$C = \frac{\varepsilon_0}{2} \mathbf{E} \cdot \nabla \times \mathbf{E} + \frac{1}{2\mu_0} \mathbf{B} \cdot \nabla \times \mathbf{B} \qquad (5)$$

When a chiral plasmonic nanoparticle is excited, both the electric and magnetic dipole modes will be induced and interact with each other. The mixed electric-magnetic dipole polarizability can be described as $\widetilde{G} = G' + iG''$. The electric dipole moment $\mathbf{p}$ and magnetic dipole moment $\mathbf{m}$ in the light field are respectively expressed as[4]

$$\widetilde{\mathbf{p}} = \widetilde{\alpha}\widetilde{\mathbf{E}} - i\widetilde{G}\widetilde{\mathbf{B}},$$
$$\widetilde{\mathbf{m}} = \widetilde{\chi}\widetilde{\mathbf{B}} + i\widetilde{G}\widetilde{\mathbf{E}}, \qquad (6)$$

where $\widetilde{\alpha} = \alpha' + i\alpha''$ is the electric polarizability, $\widetilde{\chi} = \chi' + i\chi''$ is the magnetic susceptibility, and $\mathbf{E}$ and $\mathbf{B}$ are the local fields at the nanoparticle.

The rate of excitation of the nanoparticle is

$$A^{\pm} = \langle \mathbf{E} \cdot \dot{\mathbf{p}} + \mathbf{B} \cdot \dot{\mathbf{m}} \rangle = \frac{\omega}{2} \text{Im}(\widetilde{\mathbf{E}}^* \cdot \widetilde{\mathbf{p}} + \widetilde{\mathbf{B}}^* \cdot \widetilde{\mathbf{m}}), \qquad (7)$$

where the brackets indicate an average over time. After substituting the L-G beam (Eqs. 1a-c) into the expression and considering that the particle is very thin in the z-direction, using the approximation of $z = 0$ yields the following:

$$A^{\pm} = \frac{2^{|l|}e^{-2\frac{x^2+y^2}{w_0^2}}|x^2+y^2|}{|l|!|w_0|^2} \cdot \left(\frac{\omega}{2}(\alpha''|E_0|^2 + \chi''|B_0|^2) \pm G''^{\pm}\omega\text{Im}(\widetilde{\mathbf{E}}_0^{\pm*} \cdot \widetilde{\mathbf{B}}_0^{\pm})\right). \qquad (8)$$

The difference $\Delta A = A^+ - A^-$ can be expressed as

$$\Delta A = \frac{2^{|l|}e^{-2\frac{x^2+y^2}{w_0^2}}|x^2+y^2|}{|l|!|w_0|^2} \cdot (G''^+C^+ - G''^-C^-), \qquad (9)$$

$$C = -\frac{\varepsilon_0\omega}{2}\text{Im}(\widetilde{\mathbf{E}}_0^* \cdot \widetilde{\mathbf{B}}_0) \qquad (10)$$

where C is superchiral field, which expresses the chirality of the electromagnetic field. For circularly polarized plane wave, $C_{CPL} = \pm\frac{\varepsilon_0\omega}{2c}|\widetilde{\mathbf{E}}_0|^2$ (+ refers to RCP, − referes to LCP) and for linear plane wave, $C = 0$. $\widetilde{\mathbf{E}}_0$ and $\widetilde{\mathbf{B}}_0$ here refer to the same expression as CPL plane wave. From the expression it is clear that the vortex beam only adds a spatial distribution coefficient related to $|l|$ in the chiral field as compared with an ordinary beam. If the vortex beam is a plane wave with $l = 0$, the CD reduces to[28]

$$\Delta A = G''^+C^+ - G''^-C^-. \qquad (11)$$

It can therefore be deduced that the OAM does not affect the sign of the chiral field at the waist on the paraxial approximation, and only affects the spatial distributions (Figure 4a). The difference between the resonant peaks and dichroisms can be discussed from the aspects of the mixed polarizability and field, as shown in Eq. (11). As is well known, a LG beam possesses zero amplitude at the axis, and there is a small hole with a very weak field in the middle of the beam, the size of which depends on $|l|$.[48] During excitation, the nanoparticle is in the middle of the spot and the size of the particle is comparable with that of the hole. Therefore, the particle will experience a stronger field on the edges for $|l| \neq 0$ beams and will be excited on the edges with retarded phase at different time points.[49] The polarization also varies with time for $|s| \neq 0$. For example, for $l = +1, s = +1$, the vortex illumination isophase surface will sweep over the particle in an anticlockwise direction with time, and the polarization at each point will rotate anticlockwise as well, as shown in Figure 4b. However, for $l = +1, s = -1$, the vortex illumination isophase surface will sweep over the particle in the anticlockwise direction with time, but the polarization at each point will rotate clockwise (Figure 4b). Other combinations like $l = \pm 2, s = \pm 1$ are analogous. The combination will cause a periodic time series, and the particle will be excited in different parts with different polarizations, as shown in Figure 4b. Therefore, multiplex modes are excited on the particle, as shown in Figures 4d and 4e.

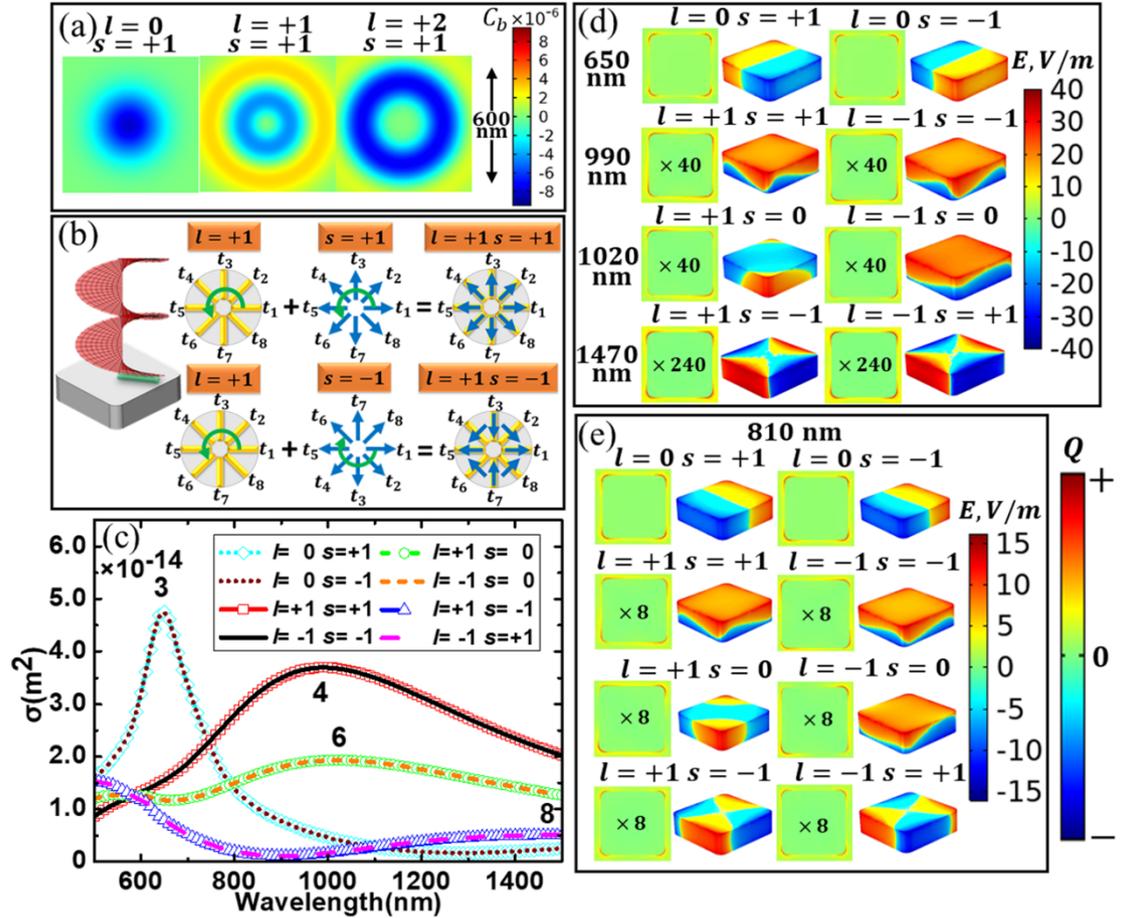

Figure 4. (a) Superchiral field of the background electric field $C_b$ when $l = 0, +1, +2$ and $s = +1$. (b) The combination of $l = +1$ and $s = +1$ in a periodic time series. The time sequence refers to the time line in one period. The yellow radial bars in the first column indicate the wavefront of the vortex beam arrives at the particle at that time. The blue arrows in the second column indicate the polarization at that time. The third column is the combination of the two (OAM and SAM). The green arrows indicate the time sequence. (c) Extinction cross-section spectra of an achiral 4-fold rotational symmetrical structure under incident light with $l = 0, \pm 1$ and $s = 0, \pm 1$. (d, e) Corresponding electric field intensity (column 1 and 3 for each subpanel) and surface charge distribution (column 2 and 4 for each subpanel) at the selected special location.

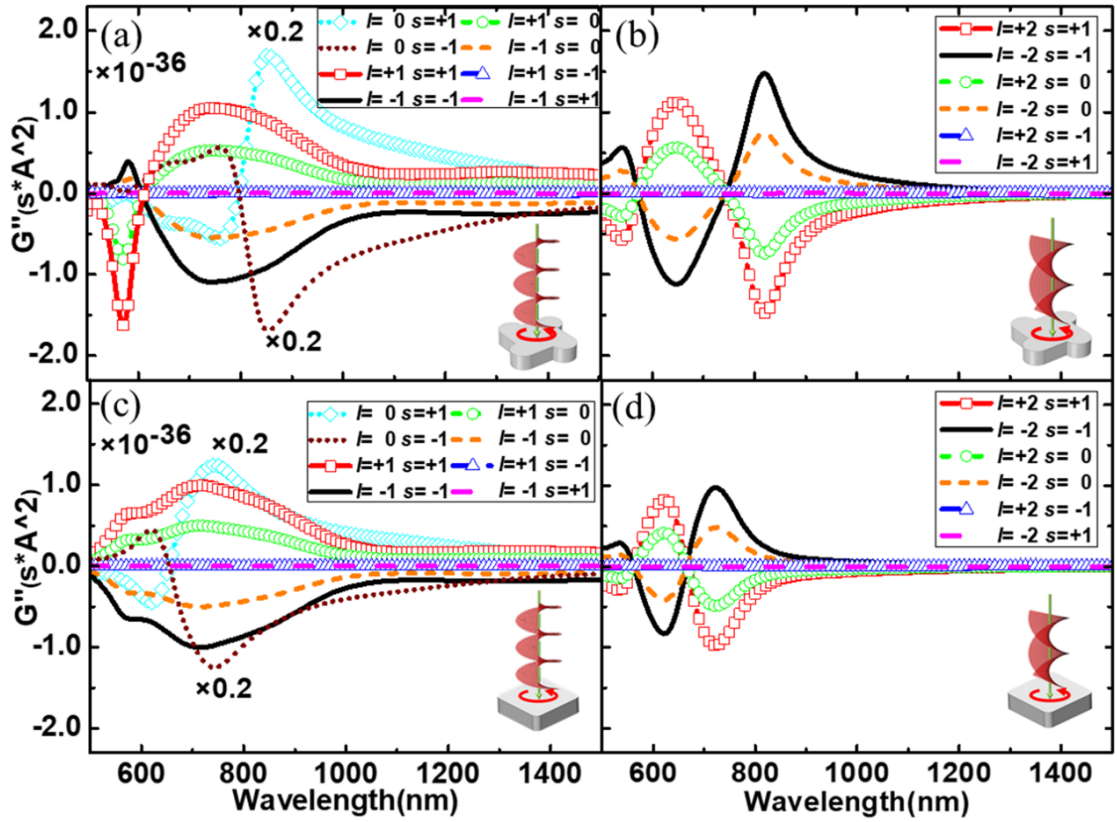

Figure 5. (a, b) $G''$ value (normalized with volume) of a gold cuboid-protuberance chiral structure when $l = 0, \pm1, \pm2$. (c, d) $G''$ value of an achiral 4-fold rotational symmetrical structure when $l = 0, \pm1, \pm2$. The value of $l = 0$ was multiplied by 0.2 for visibility in the graph.

The second aspect of the dichroism is related to the mixed polarization $G'' \propto \text{Im}(\mathbf{p}^* \otimes \mathbf{m}) \approx G_{x'x'} + G_{y'y'} + G_{z'z'} = \text{Im}(\mathbf{p}^* \cdot \mathbf{m})$ due to the interaction of the electric and magnetic dipole modes. The electric and magnetic dipole moments are respectively given as[28] $\mathbf{p} = \int d^3 r' \mathbf{r}' \rho(\mathbf{r}')$ and $\mathbf{m} = \frac{1}{2}\int d^3 r' (\mathbf{r}' \times \mathbf{J})$, where $\rho(\mathbf{r}')$ is the charge density and $\mathbf{J}(\mathbf{r}')$ is the current density. In our calculation, the optical density of states are already included in $\mathbf{p}$ and $\mathbf{m}$, but the dimension is not included. So the dimension of $G''$ should be $s * A^2 * eV^{-1}$. Different from the chiral molecules, the mixed electric-magnetic polarizability of plasmonic structures varies with different incident fields because the light field will induce different modes and affect the interaction between the electric and magnetic modes. The vortex light beam has a reshape wave front compared with the plane wave for the incident beam field, and the partial illumination of the particle by the reshaped beam will induce effective electric and magnetic oscillation modes of the particle, thus inducing different polarizabilities and $G''$ values. Figure 5 indicates that, despite whether the structure is chiral or achiral, there will be an induced $G''$ that varies with $l$. Additionally for the vortex beam ($|l| \neq 0$), the value of $G''$ decreases as $|j|$ decreases for a fixed value of $|l|$. The $G''$ values for different $|l|$ values are not comparable here because the excitation field intensity of the illumination on the particle is varied.

According to Eq. (11), the $G''$ values here are consistent with the dichroism presented in Figure 2b. The analysis here provides a straightforward and fundamental representation of the vortex dichroism.

To further investigate the dichroism, the combination of OAM and SAM beams with $l = \pm 2$ was simulated, as shown in Figure 6. Figure 6a indicates that the reshaped beam induced new resonant peaks. When $l = \pm 2$, the resonant peak of the chiral structure splits at 810 nm as compared with that when $l = \pm 1$, which may be due to the enhancement of the interaction between odd plasmon modes caused by the change of the $l$ value (as shown in Figure 6a); the beam illuminates on the structure in two parts (Figure 6b insets). Figure 6b shows the CD; compared with the previous dichroism, the $\Delta \sigma$ values of the two structures at $l = \pm 2$ are about one order of magnitude smaller than those at $l = \pm 1$. It is believed that the interaction between the structure and the vortex light at $l = \pm 2$ is weaker than that at $l = \pm 1$ due to the increase of the radius of the donut-shaped field, which leads to blue-shifting and smaller values of $G''$ and $C$ on the particle.

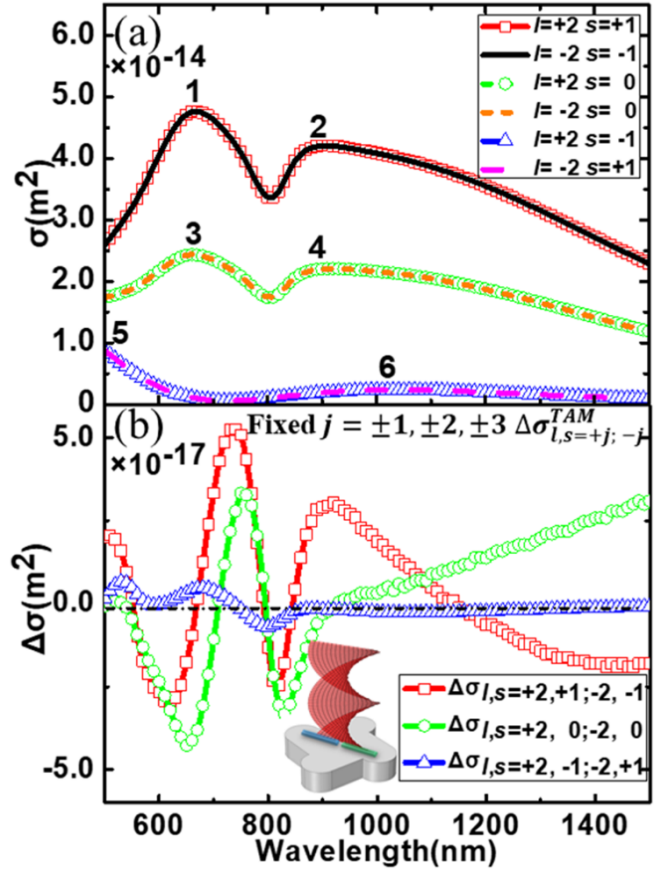

Figure 6. (a) Extinction cross-section spectra of a gold cuboid-protuberance chiral structure under incident light with $l = \pm 2$ and $s = \pm 1$. (b) Corresponding three dichroisms.

A more detailed $\Delta \sigma$ with $l = \pm 2$ is presented in Figure 7, and is similar to that at $l = \pm 1$. In Figures 7a-b, the dichroisms of $\Delta \sigma_{l,s=l,+1;\, l,-1}$ when $l = \pm 2$ and $\Delta \sigma_{l,s=+2,s;\, -2,s}$ when $s = 0, \pm 1$ are plotted. When $l = 0$ or $s = 0$, the magnitude of the dichroism is very small as compared with that when $l = \pm 2$ or $s = \pm 1$. As shown in Figures 7a-b, the symmetry between $\Delta\sigma_{l,s=+2,+1;\ +2,-1}$ and $\Delta\sigma_{l,s=+2,+1;\ -2,+1}$ regarding the spectral pattern is obvious. More combinations of $l$ and $s$ are provided in Figure 7c-f. The conclusion regarding the value of $j$ for $l = \pm 1$ is still valid here. It is worthwhile to recognize that the conclusion (the decrease of the main CD peak with the decrease of $|j|$) is likely valid for different values of $|l|$ if the excitation fields in the middle spot maintain the same intensity.

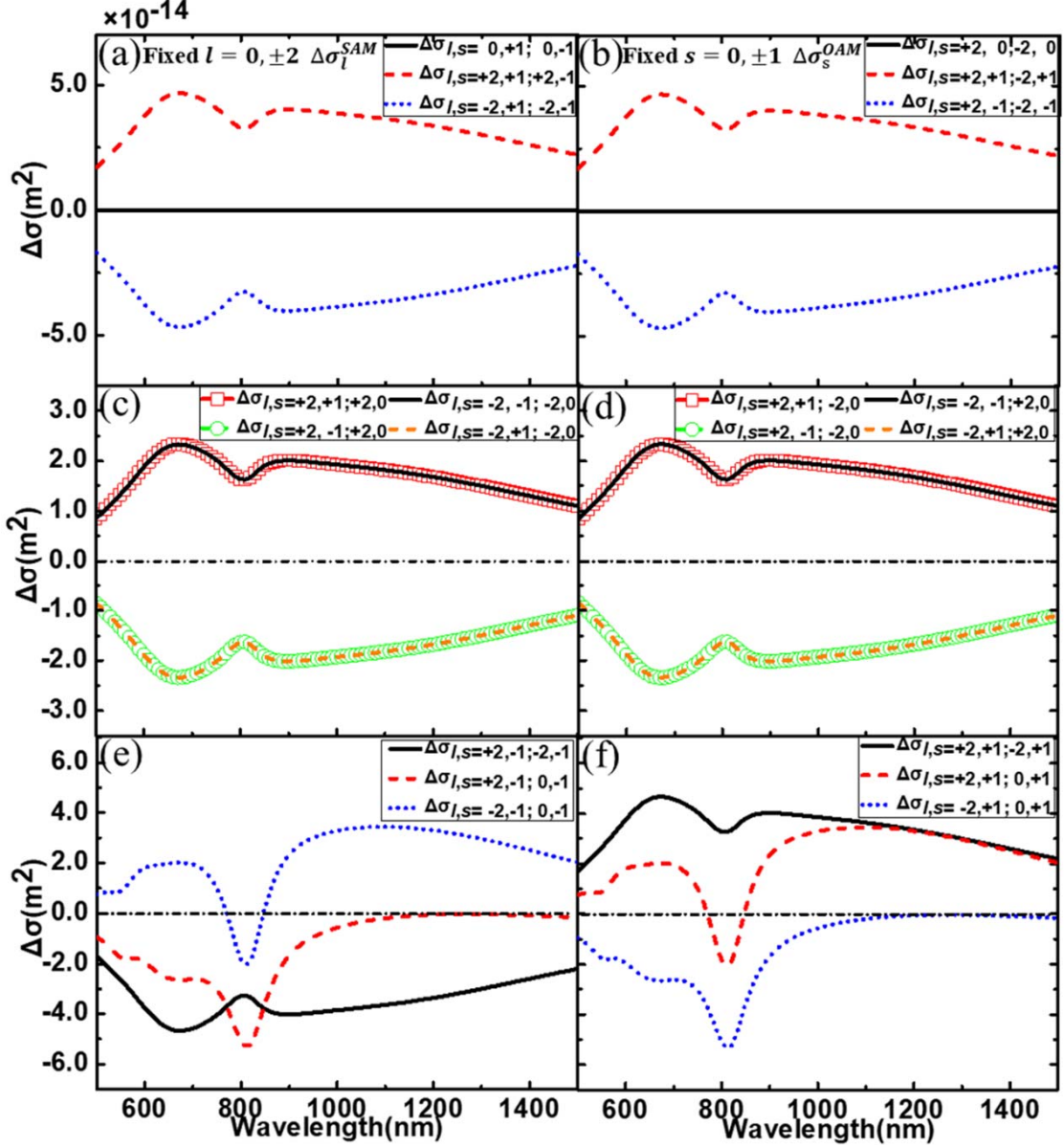

Figure 7. (a) The dichroism of $\Delta\sigma_{l,s=l,+1;\ l,-1}$ when $l = 0, \pm 2$. (b) The dichroism of $\Delta\sigma_{l,s=+2,s;\ -2,s}$ when $s = 0, \pm 1$. (c) The dichroism of $\Delta\sigma_{l,s=l,s;\ l,0}$ when $l = \pm 2$, $s = \pm 1$. (d) The dichroism of $\Delta\sigma_{l,s=l,s;\ -l,0}$ when $l = \pm 2$, $s = \pm 1$. (e-f) The dichroism of $\Delta\sigma_{l,s=+2,s;\ -2,s}$, $\Delta\sigma_{l,s=+2,s;\ 0,s}$, and $\Delta\sigma_{l,s=-2,s;\ 0,s}$ when $s = \pm 1$.

**Superchiral field**

In addition to the dichroism, the OAM-reshaped vortex beam can possibly affect the entire superchiral field $C$ defined in Eq. 10, which expresses the chirality of the electromagnetic field. Regarding the background chiral field included in Eq. (10), only the distributions are changed for the incident beam with different values of $l$ (Figure 8a), which is as expected. However, because of the illumination of light wavefront on the particle in time sequence with different polarization-induced modes, the total fields yielded by the nanoparticle under opposite values of $l$ are very different, as shown in Figure 8b. It can be determined from the figure that when $l$ and $s$ have the same sign, the superchiral field will be enhanced as compared with the case in which $l = 0$, and will be much stronger than in the cases in which $l$ and $s$ have opposite signs. For the case in which $l = +2$, the superchiral field is weak because the radius of the donut-shaped illumination field becomes so large that the incident intensity on the particle itself is weak. Therefore, when $l$ and $s$ have opposite signs, they will cancel the superchiral field. The chiral field enhancements (as compared with CPL in a plane wave with $C_{CP} = \pm \varepsilon_0/2c\omega E_0^2$) are shown in Figure 8c, and the phenomenon is similar to that presented in Figure 8b. From the enhancement distributions, it is evident that, for the increase of $|j|$ when the value of $|l|$ is fixed, the maximum superchiral field enhancement increases substantially. The results indicate that, in addition to SAM, OAM can also be used to enhance the chiral field.

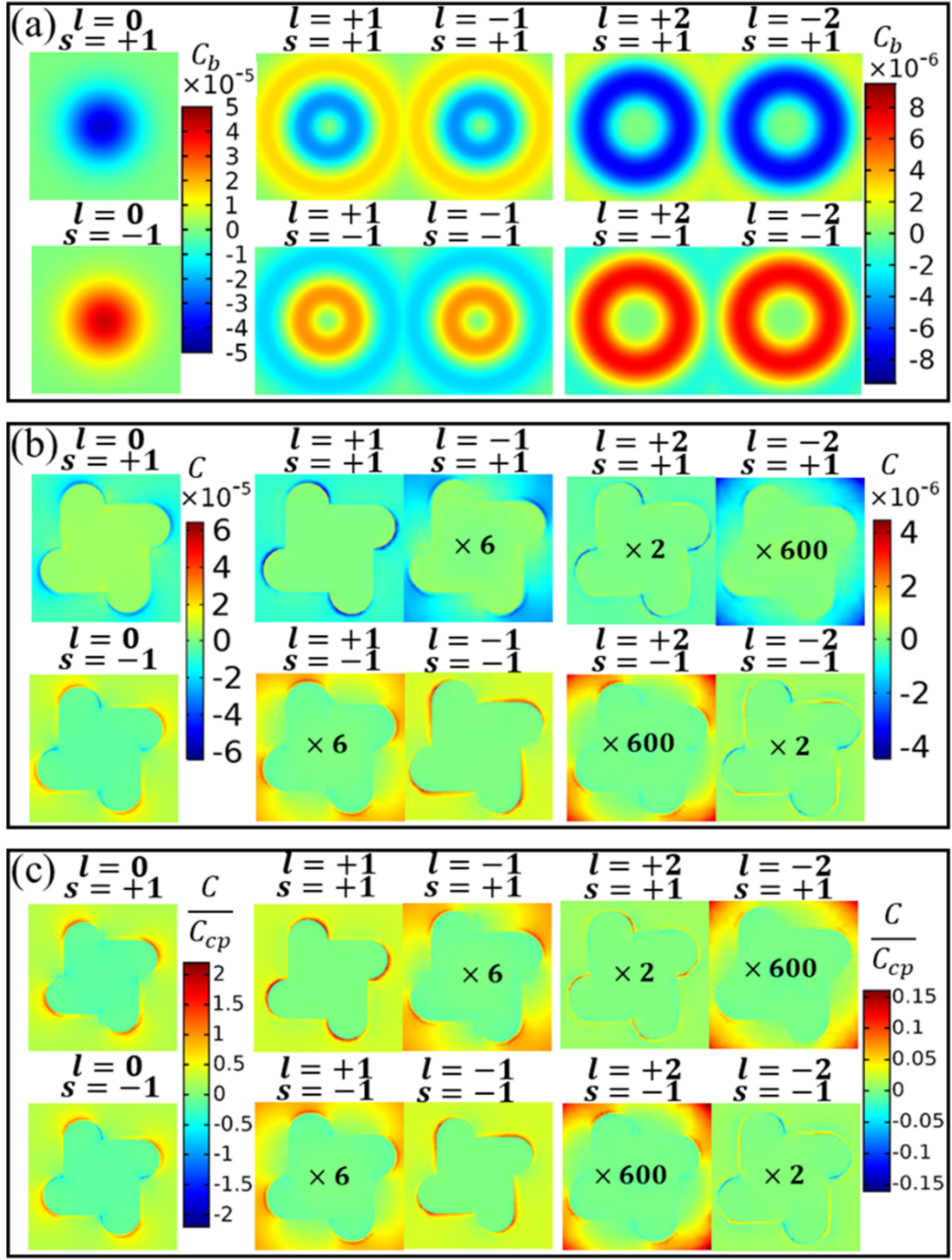

Figure 8. (a) Top view of optical chirality under a background electric field at 1000 nm when $l = 0, \pm 1, \pm 2$ and $s = \pm 1$. (b) Optical chirality of a gold cuboid-protuberance chiral structure under the same conditions. (c) Optical chirality enhancement factor[22] $C/C_{CP}$ under the same conditions, where $C_{CP} = \pm \varepsilon_0/2c\omega E_0^2$.

**Emission angle**

As the OAM excites different parts of the nanoparticle, which leads to the emergence of more modes, the emission angle of the scattered light is expected to be different because of the interference of the modes emission in far fields; in particular, the interacting electric and magnetic dipoles act as a Huygens source. The emission angle can usually be detected by the Fourier back plane (FBP) imaging technique.[50] The FBP pattern is got from the interference of all of the discrete-mesh dipoles emission in far field.[51-53]

The FBP imaging results for the particle at specific peaks are plotted in Figure 9. As shown in Figure 9a when $l = 0$ and the $s$ value changes, the FBP reveals a symmetrical distribution; when $s = 0$ and the $l$ value changes, the FBP does not change significantly; when $l$ and $s$ are not equal to 0, the FBP presents a symmetrical distribution and is not the same as the previous distribution, which demonstrates that $s$ is related to the symmetry of the FBP. Additionally, although $l$ is irrelevant, it has an influence on the distribution of the FBP. Therefore, OAM can also be used to multiply the light emission directions. The emission angle also serves as proof that the coupled electric and magnetic dipole modes are quite different because each electric and magnetic dipole combination can be treated as a Huygens source that has a uniform directed emission, and can extend the electric-magnetic interaction mechanism of a vortex with OAM.

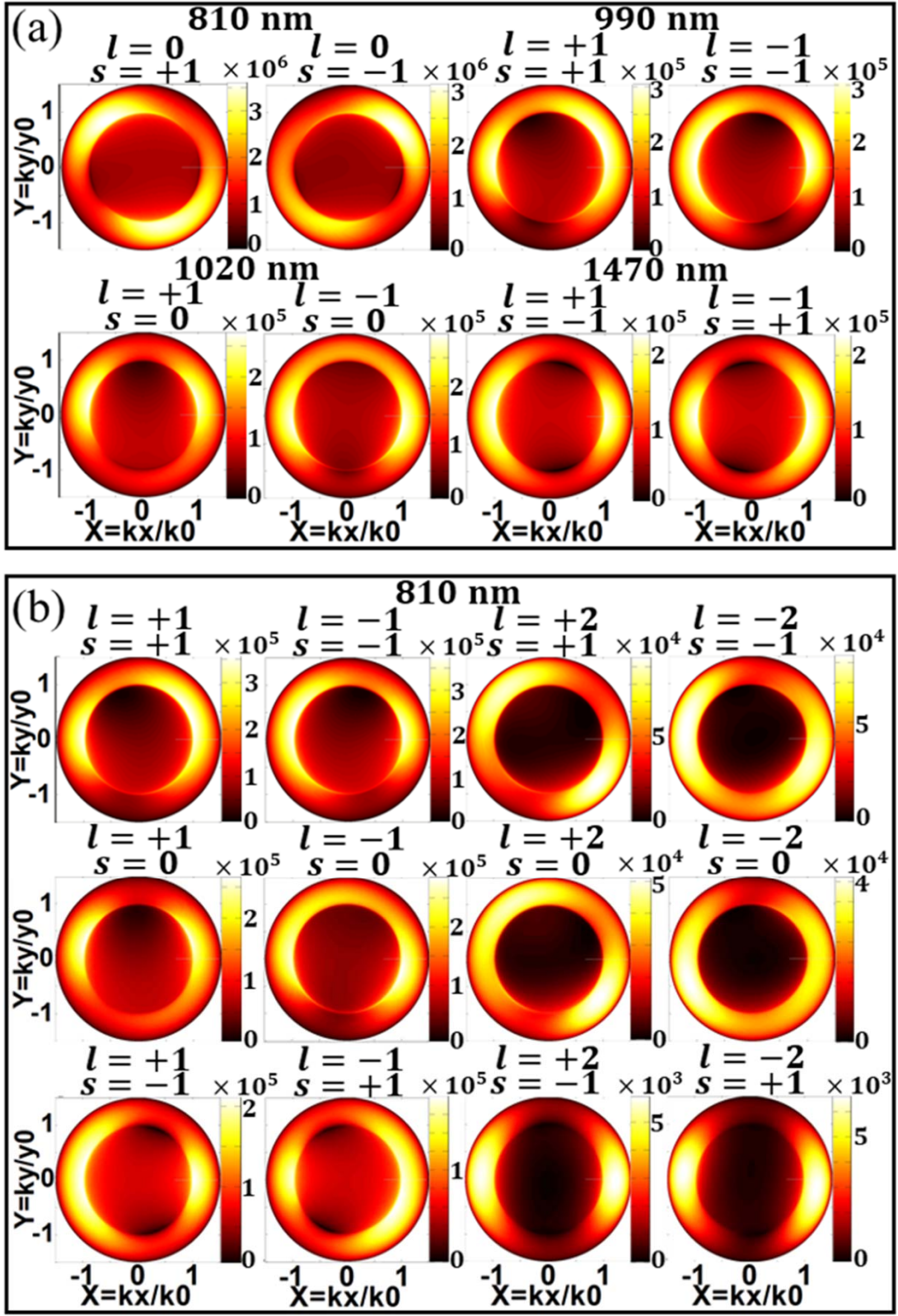

Figure 9. (a) FBP imaging of a gold cuboid-protuberance chiral structure at the peaks at 810, 990, 1020, and 1470 nm with $l = 0, \pm 1$ and $s = \pm 1$. (b) FBP imaging at 810 nm when $l = 0, \pm 1, \pm 2$ and $s = \pm 1$.
16

## Discussion and conclusions

In this paper, the dichroism responses of a chiral cuboid-protuberance structure excited by an L-G beam with both OAM and SAM were investigated. The combination of the polarization variation of SAM and the spatial variation of OAM caused rich, excited modes that exhibited quite different resonant peaks. When $|l| \neq 0$ and $|s| \neq 0$, the dichroism had a very large value and decreased with the decrease of $|j|$. In presence of nonzero $|l|$, the CP-induced ($s = \pm 1$) dichroism is greatly enhanced against the case of $l = 0$, as well as in presence of non-zero s, the OAM-induced ($l = \pm 1$) dichroism is enhanced against the case of $s = 0$. The mixed electric-magnetic polarization $G''$, which reflects the interaction of the induced electric and magnetic modes, was attributed to the response of the large dichroism. Because of the wavefront-reshaping effect of the vortex beam, the particle was found to be excited with the vortex wavefront in a time series, which induces more plasmonic modes on the particle and ultimately changes the strength and interaction of the electric and magnetic dipoles. Because the intensity in the middle area of the focused beam becomes very weak as $l$ increases, the value of the spectra will be very small. To get comparable results, the beam waist is fixed, which makes the paraxial approximation not valid in the long wavelengths[48]. So in our study, the $z$ component of the excitation beam is ignored (because the nanoparticle is very thin as that the polarizability in $z$ direction is ignorable)[54], which is still quite similar to the plane wave in the waist position. Because the phenomenon in this work is caused by the phase retardation of vortex beam in time, our approximation still works for the mechanism. And additional simulations with paraxial condition show similar phenomenon. The co-action of OAM and SAM was also found to significantly change the superchiral near-field distributions, which provides a new approach for the manipulation of light on the nano-scale. The combination of OAM and SAM would be expected to bring rich phenomenon in cases of light interaction with metamaterials formed by combination of multiple identical such chiral elements. Therefore, this study paves the way for the investigation and application of vortex beams with OAM and SAM in chirality.

## Acknowledgments

## Funding

This research was supported by the National Natural Science Foundation of China (Grant Nos. 11704058, 11974069), the National Special Support Program for High-level Personnel Recruitment (Grant No. W03020231), the LiaoNing Revitalization Talents Program (Grant No. XLYC1902113), the Program for Liaoning Innovation Research Team in University (Grant No. LT2016011), the Science and Technology Foundation of Dalian (Grant No. 2017RD12), and the Fundamental Research Funds for the Central Universities (Grant No. DUT19RC(3)007).

## Conflicts of interest

The authors declare no competing financial interest.

# References

[1]     A. O. Govorov, Z. Fan, P. Hernandez, J. M. Slocik, and R. R. Naik, Nano Letters **10**, 1374 (2010).

[2]     A. O. Govorov and Z. Fan, Chemphyschem **13**, 2551 (2012).

[3]     L. Hu, F. Xi, L. Qv, and Y. Fang, ACS Omega **3**, 1170 (2018).

[4]     Y. Tang and A. E. Cohen, Physical Review Letters **104**, 163901, 163901 (2010).

[5]     Y. Tang and A. E. Cohen, Science **332**, 333 (2011).

[6]     X. Ren, W. Lin, Y. Fang, F. Ma, and J. Wang, RSC Advances **7**, 34376 (2017).

[7]     Y. Tang, Y. Huang, L. Qv, and Y. Fang, Nanoscale Res Lett **13**, 194 (2018).

[8]     M. Liu, T. Zentgraf, Y. Liu, G. Bartal, and X. Zhang, Nature Nanotechnology **5**, 570 (2010).

[9]     C. Jack, A. S. Karimullah, R. Leyman, R. Tullius, V. M. Rotello, G. Cooke, N. Gadegaard, L. D. Barron, and M. Kadodwala, Nano Letters **16**, 5806 (2016).

[10]    R. Zhao, L. Zhang, J. Zhou, T. Koschny, and C. M. Soukoulis, Physical Review B **83**, 035105 (2011).

[11]    R. Schreiber *et al.*, Nature Communications **4**, 2948 (2013).

[12]    P. Kuhler, E. M. Roller, R. Schreiber, T. Liedl, T. Lohmuller, and J. Feldmann, Nano Lett **14**, 2914 (2014).

[13]    X. Lan, X. Zhou, L. A. McCarthy, A. O. Govorov, Y. Liu, and S. Link, J Am Chem Soc **141**, 19336 (2019).

[14]    M. Hentschel, V. E. Ferry, and A. P. Alivisatos, Acs Photonics **2**, 1253 (2015).

[15]    X. Yin, M. Schaferling, B. Metzger, and H. Giessen, Nano Lett **13**, 6238 (2013).

[16] S. Droulias and V. Yannopapas, The Journal of Physical Chemistry C **117**, 1130 (2013).

[17] Z. Fan and A. O. Govorov, Journal of Physical Chemistry C **115**, 13254 (2011).

[18] M. Schäferling, X. Yin, N. Engheta, and H. Giessen, ACS Photonics **1**, 530 (2014).

[19] M. Esposito, V. Tasco, F. Todisco, M. Cuscuna, A. Benedetti, D. Sanvitto, and A. Passaseo, Nat Commun **6**, 6484 (2015).

[20] M. Esposito, V. Tasco, F. Todisco, A. Benedetti, I. Tarantini, M. Cuscuna, L. Dominici, M. De Giorgi, and A. Passaseo, Nanoscale **7**, 18081 (2015).

[21] S. Zu, Y. Bao, and Z. Fang, Nanoscale **8**, 3900 (2016).

[22] R. Ogier, Y. Fang, M. Svedendahl, P. Johansson, and M. Käll, ACS Photonics **1**, 1074 (2014).

[23] J. A. Fan *et al.*, Science **328**, 1135 (2010).

[24] Z. Fan and A. O. Govorov, Nano Lett **12**, 3283 (2012).

[25] M. Schaeferling, D. Dregely, M. Hentschel, and H. Giessen, Physical Review X **2**, 031010, 031010 (2012).

[26] Y. Luo, C. Chi, M. Jiang, R. Li, S. Zu, Y. Li, and Z. Fang, Advanced Optical Materials **5**, 1700040 (2017).

[27] X. Tian, Y. Fang, and M. Sun, Sci Rep **5**, 17534 (2015).

[28] L. Hu, X. Tian, Y. Huang, L. Fang, and Y. Fang, Nanoscale **8**, 3720 (2016).

[29] L. Hu, Y. Huang, L. Pan, and Y. Fang, Sci Rep **7**, 11151 (2017).

[30] L. Allen, M. W. Beijersbergen, R. J. Spreeuw, and J. P. Woerdman, Phys Rev A **45**, 8185 (1992).

[31] A. A. Robert C. Devlin, Noah A. Rubin, J. P. Balthasar Mueller, Federico Capasso, Science

**358**, 896 (2017).

[32] H. Ren *et al.*, Nat Commun **10**, 2986 (2019).

[33] Z. Liu, S. Yan, H. Liu, and X. Chen, Physical Review Letters **123**, 183902 (2019).

[34] S. Franke-Arnold, Philos Trans A Math Phys Eng Sci **375** (2017).

[35] E. Nagali, F. Sciarrino, F. De Martini, L. Marrucci, B. Piccirillo, E. Karimi, and E. Santamato, Phys Rev Lett **103**, 013601 (2009).

[36] D. S. Ding, W. Zhang, Z. Y. Zhou, S. Shi, G. Y. Xiang, X. S. Wang, Y. K. Jiang, B. S. Shi, and G. C. Guo, Phys Rev Lett **114**, 050502 (2015).

[37] S. W. H. a. J. Wichmann, OPTICS LETTERS **19**, 780 (1994).

[38] D. Pan, H. Wei, L. Gao, and H. Xu, Phys Rev Lett **117**, 166803 (2016).

[39] F. C. S. Martin P. J. Lavery, Stephen M. Barnett, Miles J. Padgett, SCIENCE **341**, 537 (2013).

[40] D. G. Grier, Nature **424**, 810 (2003).

[41] K. A. Forbes, Phys Rev Lett **122**, 103201 (2019).

[42] M. Mirhosseini, O. S. Magaña-Loaiza, M. N. O'Sullivan, B. Rodenburg, M. Malik, M. P. J. Lavery, M. J. Padgett, D. J. Gauthier, and R. W. Boyd, New Journal of Physics **17**, 033033 (2015).

[43] A. E. Willner *et al.*, Advances in Optics and Photonics **7** (2015).

[44] N. Bozinovic, Y. Yue, Y. Ren, M. Tur, P. Kristensen, H. Huang, A. E. Willner, and S. Ramachandran, Science **340**, 1545 (2013).

[45] R. M. Kerber, J. M. Fitzgerald, D. E. Reiter, S. S. Oh, and O. Hess, ACS Photonics **4**, 891 (2017).

[46] R. M. Kerber, J. M. Fitzgerald, X. Xiao, S. S. Oh, S. A. Maier, V. Giannini, and D. E. Reiter, New Journal of Physics **20** (2018).

[47] R. M. Kerber, J. M. Fitzgerald, S. S. Oh, D. E. Reiter, and O. Hess, Communications Physics **1** (2018).

[48] A. Bekshaev, M.Soskin, and M.Vasnetsov, Paraxial Light Beams with Angular Momentum, New York, Nova Science Publishers (2008).

[49] A. Bekshaev, K. Y. Bliokh, and M. Soskin, Journal of Optics **13**, 053001 (2011).

[50] Y. Fang, Physics Experimentation **39**, 1 (2019).

[51] W. Lukosz, J. Opt. Soc. Am. **69**, 1495 (1979).

[52] M. A. Lieb, J. M. Zavislan, and L. Novotny, J. Opt. Soc. Am. B **21**, 1210 (2004).

[53] T. Shegai, V. D. Miljković, K. Bao, H. Xu, P. Nordlander, P. Johansson, and M. Käll, Nano Letters **11**, 706 (2011).

[54] A. Mendoza-Galván, K. Järrendahl, A. Dmitriev, T. Pakizeh, M. Käll, and H. Arwin, Optics Express **19**, 12093 (2011).